\documentclass[aip,amsmath,amssymb,reprint]{revtex4-1}

\usepackage{bm}  
\usepackage{graphicx}
\usepackage{dcolumn}

\usepackage[utf8]{inputenc}
\usepackage[T1]{fontenc}
\usepackage{mathptmx}

\begin{document}

\preprint{AIP/prova}

\title{On the Fluctuation-Dissipation Relation in non-equilibrium and non-Hamiltonian systems}

\author{A. Sarracino}
\email{alessandro.sarracino@unicampania.it}
\affiliation{Dipartimento di Ingegneria, Universit\`a della Campania ``L. Vanvitelli'', via Roma 29, 81031 Aversa (CE), Italy}

\author{A. Vulpiani}
\email{angelo.vulpiani@roma1.infn.it}
\affiliation{Dipartimento di Fisica, Universit\`a Sapienza - p.le A. Moro 2, 00185 Roma, Italy}
\affiliation{Centro Interdisciplinare ``B. Segre'', Accademia dei Lincei, 00165 Roma, Italy}

\date{\today}

\begin{abstract}
We review generalized Fluctuation-Dissipation Relations which are
valid under general conditions even in ``non-standard systems'', e.g.
out of equilibrium and/or without a Hamiltonian structure.  The
response functions can be expressed in terms of suitable correlation
functions computed in the unperperturbed dynamics.  In these
relations, typically one has nontrivial contributions due to the form
of the stationary probability distribution; such terms take into
account the interaction among the relevant degrees of freedom in the
system.  We illustrate the general formalism with some examples in
non-standard cases, including driven granular media, systems with a
multiscale structure, active matter and systems showing anomalous
diffusion.
\end{abstract}

\maketitle 

\begin{quotation}
  The Fluctuation-Dissipation Theorem is a central result in
  equilibrium statistical mechanics.  It allows one to express the
  linear response of a system to an external perturbation in terms of
  the spontaneous correlations, and its derivation is based on the
  detailed balance condition.  In the last decades a great effort has
  been devoted to generalize this fundamental result to systems where
  detailed balance does not hold, because of the presence of some
  forms of dissipation, or energy and particle currents, that induce
  non-equilibrium conditions. Among the huge variety of systems
  belonging to this class, let us mention active and biological
  matter, driven granular media, molecular motors, and slow relaxing
  glasses. The derivation of a generalized Fluctuation-Dissipation
  Relation, and the investigation of its peculiar features in
  non-standard systems play therefore a central role in the building of
  a general theory of statistical mechanics beyond equilibrium.
\end{quotation}

\section{Introduction}
\label{Introduction}

One of the general results of the near-equilibrium statistical
mechanics is the existence of a precise relation between the
spontaneous fluctuations of the system and the response to external
perturbations of physical observables. This result allows for the
possibility of studying the response to time-dependent external
fields, by analyzing time-dependent correlations. The idea dates back
to Einstein's work on Brownian motion~\cite{E05}. Later,
Onsager~\cite{O31,O31b} stated his regression hypothesis according to
which the relaxation of a macroscopic perturbation follows the same
laws governing the dynamics of fluctuations in equilibrium systems.

The Fluctuation-Dissipation Relation (FDR) theory was initially
obtained for Hamiltonian systems near thermodynamic equilibrium, but
now it is known that a generalized FDR holds for a broader class of
systems.  Among these, the FDR has been widely investigated in the
context of turbulence (and more generally statistical fluid
mechanics): let us mention for instance the seminal work by
Kraichnann~\cite{Kr00,Kr59}, and his attempt to obtain a closure
theory from an assumption on the FDR.  In recent years, the FDR
attracted the interest of the scientific community active in the
modelling of geophysical systems, in particular for climate
dynamics~\cite{L75,B80,NBH93,MW06,AM08,N1}.  Moreover, another wide
field of research where FDR plays a central role is the stochastic
thermodynamics theory~\cite{Seifert}, including models for colloidal
systems, granular and active matter.

Here we present a brief review of some recent results on the theory of
FDR, with specific focus on non-standard situations: systems with
negative temperature, systems with many degrees of freedom and a
multiscale structure, or systems with anomalous transport
properties. We discuss in particular some subtle, non-trivial aspects
related to these peculiar cases. Our work represents an update of the
previous review~\cite{capklages}.  The interested reader can find more
exhaustive reviews of the general subject
in~\cite{CR03,BPRV08,cuglirev,Seifert,maes3,PSV17}. For a more
rigorous discussion see also~\cite{baladi}.

The paper is organized as follows. In Section~\ref{GFDR} we present
two generalized FDRs valid also in non-equilibrium systems and discuss
some subtle points. The first relation involves the probability
distribution function (PDF) of the stationary state, while the second
one includes a quantity which depends on the transition rates of the
model. In Section~\ref{real} we illustrate some examples where FDR are
applied in non-standard cases, such as anomalous diffusion, systems
with different time scales, and some models of active matter.  In
Section~\ref{conc} we draw some conclusions and mention perspectives
for future work.

\section{Generalized Fluctuation-Dissipation Relation}
\label{GFDR}

The FDR theory was initially developed within the context of
equilibrium statistical mechanics of Hamiltonian systems.  This led to
misleading claims on the (supposed) limited validity of the
FDR~\cite{RS78}.  Indeed, in the following we will see that it is
possible to derive a generalized FDR, which holds under rather general
hypotheses, also in non-Hamiltonian systems~\cite{DH75,FIV90,BLMV03}.

\subsection{van Kampen's objection to the FDR}

Let us briefly discuss an objection by van Kampen~\cite{vK71} to the
original (perturbative) derivation of the FDR.  From a technical point
of view such a criticism can be rejected: however it had the merit
of stimulating a deeper understanding of the FDR and its validity range.

In the dynamical systems terminology, van Kampen's
argument is the following.  Given an impulsive perturbation $\delta
{\bf x}(0)$ on the state {\bf x} of the system at time $t=0$, the difference
between the perturbed trajectory and the unperturbed one is
\begin{equation}
\label{3.1}
\delta  x_i(t)=
\sum_{j} \frac{\partial x_i(t)}{\partial x_j(0)} \delta  x_j(0)
+O(|\delta {\bf x}(0)|^2).
\end{equation}
Averaging over the initial conditions, one has the mean 
response function:
\begin{equation}
\label{3.2}
R_{i,j}(t)= \Biggl \langle \frac{\delta x_i(t)} {\delta x_j(0)}
\Biggr \rangle =\int \frac{\partial x_i(t)} {\partial x_j(0)}
\rho({\bf x}(0)) d{\bf x}(0 ) \,\,.
\end{equation}
In the case of the equilibrium statistical mechanics, since $\rho({\bf
  x}) \propto \exp\Bigl( -\beta H(\bf{x}) \Bigr)$, after an
integration by parts, one obtains
\begin{equation}
\label{3.3}
R_{i,j}(t)= \beta \Biggl \langle x_i(t)
 \frac {\partial  H({\bf x}(0))} 
{ \partial x_j(0)} \Biggr \rangle \,.
\end{equation}
It is easy to realize that the above result is nothing but the
differential form of the usual FDR.

In the presence of chaos, however, the terms ${\partial
  x_i(t)}/{\partial x_j(0)}$ grow exponentially as $e^{\lambda t}$,
where $\lambda$ is the Lyapunov exponent. Therefore the linear
expansion~(\ref{3.1}) is not accurate for a time larger than $(1/
\lambda) \ln (L/|\delta{\bf x}(0)|)$, where $L$ is the typical
fluctuation of the variable ${\bf x}$. Thus, the linear response
theory is expected to be valid only for extremely small and
nonphysical perturbations (or times).  For instance, according to this
argument, requiring that the FDR holds up to $1 s$ when applied to the
electrons in a typical conductor, would imply a perturbing electric
field smaller than $10^{-20} V/m$, in clear disagreement with the
experience.

The success of the linear theory for the computation of transport
coefficients (e.g. electric conductivity) in terms of correlation
functions of the unperturbed system, is evident, and its validity has
been, directly and indirectly, verified in a huge number of cases.
Kubo suggested that the origin of the effectiveness of the FDR theory
may reside in the ``constructive role of chaos'': ``{\it instability
  [of the trajectories] instead favors the stability of distribution
  functions, working as the cause of the mixing}''~\cite{K86}.  The
following derivation~\cite{FIV90} of a generalized FDR supports this
intuition.

\subsection{A Generalized FDR}
\label{GFDRs}

One of the most intense research fields in non-equilibrium statistical
mechanics addressed the issue of the fluctuation-dissipation theorem
when the system under study is out of equilibrium. This situation can
be due both to the presence of external forcing and continuous
dissipation, so that a stationary state is reached, and to a
very slow relaxation, leading to a non-stationary transient dynamics.
Standard examples of systems falling in the first class are vibrated
granular materials~\cite{Puglisi} or active
particles~\cite{RevModPhys.85.1143}, while to the second class there
belong, for instance, Ising spin models, or spin and structural
glasses~\cite{CR03}.

Many results have been derived in the last decades, which extend the
validity of the FDR to the non-equilibrium
realm~\cite{BPRV08,Seifert}. Of course, these results do not share the
same generality as the equilibrium FDR, and their explicit forms can
depend on the considered model. Nevertheless, these relations can have
important applications in different contexts.  In general, their
relevance relies on the possibility to obtain information on the
non-equilibrium response of the system from the study of unperturbed
fluctuations, or vice-versa, depending on the more suitable conditions
in numerical or experimental settings.

Let us start by presenting a derivation of a generalized FDR. It is
easy to understand, see below, that it is possible to derive such a
FDR also for finite perturbations, in non equilibrium and non
Hamiltonian systems, and therefore van Kampen's critique has a
marginal role.

Consider a dynamical system ${\bf x}(0) \to {\bf x}(t)=U^t {\bf x}(0)$
with states ${\bf x}$ belonging to a $N$-dimensional space.  We can
consider the case in which the time evolution is not deterministic
(e.g., stochastic differential equations).  Let us assume the
existence of an invariant probability distribution $\rho({\bf x})$,
for which some ``absolutely continuity''-type conditions are required
(see later), and the mixing character of the system (from which its
ergodicity follows); no assumption is made on the dimensionality $N$
of the system.  Our aim is to study the behaviour of one component of
${\bf x}$, say $x_i$, when the system, whose statistical features are
described by $\rho({\bf x})$, is subjected to an initial (non-random)
perturbation such that ${\bf x}(0) \to {\bf x}(0) + \Delta {\bf
  x}_{0}$.

An instantaneous kick modifies the density of the system into
$\rho'({\bf x})$, related to the invariant distribution by $\rho'
({\bf x}) = \rho ({\bf x} - \Delta {\bf x}_0)$.  Let us introduce the
probability of transition from ${\bf x}_0={\bf x}(0)$ at time $0$ to
${\bf x}$ at time $t$, $w ({\bf x}_0,0 \to {\bf x},t)$ (of course in a
deterministic system, with evolution law $ {\bf x}(t)=U^{t}{\bf
  x}(0)$, one has $w ({\bf x}_0,0 \to {\bf x},t)=\delta({\bf
  x}-U^{t}{\bf x}_{0})$, where $\delta(\cdot)$ is the Dirac delta).
We can write an expression for the mean value of the variable $x_i$,
computed with the density of the perturbed system:
\begin{equation}
\label{3.4}
\Bigl \langle x_i(t) \Bigr \rangle ' = \int\!\int x_i \rho' ({\bf
  x}_0) w ({\bf x}_0,0 \to {\bf x},t) \, d{\bf x} \, d{\bf x}_0 \; .
\end{equation}
For the mean value of $x_i$ during the unperturbed evolution one has:
\begin{equation}
\label{3.5}
\Bigl \langle x_i(t) \Bigr \rangle = 
\int\!\int x_i \rho ({\bf x}_0) 
w ({\bf x}_0,0 \to {\bf x},t) \, d{\bf x} \, d{\bf x}_0  \; .
\end{equation}
Therefore, defining $\overline{\delta x_i} =  \langle x_i \rangle' -
\langle x_i \rangle$, we have:
\begin{eqnarray}
\label{3.6}
\overline{\delta x_i} \, (t)  &=&
\int  \int x_i \;
\frac{\rho ({\bf x}_0 - \Delta {\bf x}_0) - \rho ({\bf x}_0)}
{\rho ({\bf x}_0) } \;
\rho ({\bf x}_0) w ({\bf x}_0,0 \to {\bf x},t) 
\, d{\bf x} \, d{\bf x}_0 \nonumber \\
&=& \Bigl \langle x_i(t) \;  F({\bf x}_0,\Delta {\bf x}_0) \Bigr \rangle,
\end{eqnarray}
where
\begin{equation}
\label{3.7}
F({\bf x}_0,\Delta {\bf x}_0) =
\left[ \frac{\rho ({\bf x}_0 - \Delta {\bf x}_0) - \rho ({\bf x}_0)}
{\rho ({\bf x}_0)} \right] \; .
\end{equation}
Note that the mixing property of the system is required to
guarantee the decay to zero of the time-correlation functions and,
thus, the switching off of the deviations from equilibrium.

In the case of an infinitesimal perturbation $\delta {\bf x}(0) =
(\delta x_1(0) \cdots \delta x_N(0))$, if $\rho({\bf x})$ is
non-vanishing and differentiable, the function in (\ref{3.7}) can be
expanded to first order and one obtains:
\begin{eqnarray}
\label{3.8}
\overline{\delta x_i} \, (t)  &=&
- \sum_j \Biggl
\langle x_i(t) \left. \frac{\partial \ln \rho({\bf x})}{\partial x_j} 
\right|_{t=0}  \Biggr \rangle \delta x_j(0) \nonumber \\
&\equiv&
\sum_j R_{i,j}(t) \delta x_j(0),
\end{eqnarray}
which gives the linear response  
\begin{equation}
\label{3.9}
R_{i,j}(t) = - \Biggl \langle x_i(t) \left.
 \frac{\partial \ln \rho({\bf x})} {\partial x_j} \right|_{t=0}
\Biggr  \rangle 
\end{equation} 
of the variable $x_i$ with respect to a perturbation of $x_j$.
It is easy to repeat the computation 
for a generic observable $A({\bf x})$, yielding 
$$
\overline{A(t)}=- \sum_j \langle A({\bf x}(t)) \left.\frac{\partial \ln \rho({\bf x})} {\partial x_j} \right|_{t=0} \rangle \delta x_j(0) \,.
$$

Let us note that the study of an ``impulsive'' perturbation is not a
limitation: e.g., in the linear regime from the (differential) linear
response one can understand the effect of a generic perturbation.  For
instance, consider a system ruled by the evolution law
$$
{d {\bf x} \over dt}= {\bf Q}({\bf x})
$$
and apply an infinitesimal perturbation:
${\bf Q}({\bf x}) \to  {\bf Q}({\bf x})+ \delta {\bf Q}(t)$
with $\delta {\bf Q}(t)=0$ for $t<0$. Then one has
$$
\overline{\delta x_i(t)}=
\sum_j \int_0^t R_{ij}(t-t') \delta Q_j(t') \, dt' \,.
$$

\subsection{Finite perturbations and relevance of chaos}

We note that in the above derivation of the FDR relation we never used
any approximation on the evolution of $\delta {\bf x}(t)$.  Starting
with the exact expression~(\ref{3.6}) for the response, only a
linearization on the initial time perturbed density is needed, and
this implies nothing but the smallness of the initial perturbation.

It is easy to understand  that it is possible
to derive a FDR also for finite perturbations.
One has
$$
\overline{\delta A(t)}=\langle A({\bf x}(t)) F({\bf x}(0), \Delta {\bf x}(0))
\rangle,
$$
where the $F({\bf x}(0), \Delta {\bf x}(0))$, Eq.~(\ref{3.7}), depends on the initial
perturbation $\Delta {\bf x}(0)$ and the invariant probability distribution~\cite{BLMV03}.
We  stress again  that, from the evolution
of the trajectories difference, one can define the leading Lyapunov
exponent $\lambda$ by considering the absolute values of $\delta {\bf x}(t)$: 
at small $|\delta {\bf x}(0)|$ and large enough $t$ one has
\begin{equation}
\label{3.11}
\Bigl \langle \ln |\delta {\bf x}(t)|\Bigr \rangle \simeq 
\ln |\delta {\bf x}(0)| + \lambda t \,\, .
\end{equation}
On the other hand, in the FDR one deals with averages of quantities
with sign, such as $\overline{\delta {\bf x}(t)}$.  This apparently
marginal difference is very important and allows for the possibility
to derive the FDR relation avoiding van Kampen's objection.  However
van Kampen's remark can have a practical, but not conceptual,
relevance.  For instance in the presence of chaos the computational
error on $\overline{\delta {\bf x}(t)}$ increases as $e^{\alpha
  t}/\sqrt{M}$ where $M$ is the number of perturbations and $\alpha
\simeq 2 \lambda$; such a behaviour can be easily verified, see
e.g.~\cite{FIV90,BLMV03}.

\subsection{About the invariant measure} 
\label{invariant}

Since the generalized FDR~(\ref{3.9}) explicitly involves the
invariant measure $\rho({\bf x})$, it is useful to comment on some
features that immediately derive from the properties of $\rho$.

First, we note that in Hamiltonian systems, taking the canonical
ensemble as the equilibrium distribution, one has $ \ln \rho= -\beta
H({\bf Q},{\bf P})\, + const.$ Recalling Eq.~\eqref{3.9}, if we
denote by $x_i$ the component $q_k$ of the position vector ${\bf Q}$
and by $x_j$ the corresponding component $p_k$ of the momentum ${\bf
  P}$, from Hamilton's equations ($dq_k/dt=\partial H/\partial p_k$)
one immediately has the differential form of the usual FDR~\cite{K66,K86}
\begin{equation}
\label{3.12}
\frac{ \overline{\delta q_k} \, (t)} {\delta p_k(0)}
=\beta \Biggl \langle q_k(t) \frac {dq_k(0)} {dt} \Biggr \rangle =
- \beta \frac{d}{dt} \Bigl \langle q_k(t) q_k(0) \Bigr \rangle \, ,
\end{equation}
where time-translational invariance has been used.

\subsubsection{Systems with negative temperature}

In the most common Hamiltonian systems one has
$$
H=\sum_n {p_n^2 \over 2 m} + V(q_1, ... , q_N) \,.
$$
However, there are cases where the ``kinetic term'' is not
parabolic and therefore the previous Hamiltonian is replaced by the more general form
$$
H=\sum_n  K(p_n) + V(q_1, ... , q_N) \,.
$$
If $\{ q_n \}$ and $\{ p_n \}$ are bounded variables, for a suitable form
of $K(p)$ one can have negative absolute temperature, i,e.
for some values of the energy $E$, $\partial S(E)/\partial E$ is negative,
being $S(E)$ the microcanonical entropy~\cite{PSV17}.
Such systems are not mere curiosity,
being rather relevant for instance in hydrodynamics
and plasma physics~\cite{Joyce}.
In addition, recent experimental measurements  showed 
the presence of a negative absolute temperature
in cold atoms systems~\cite{Braun}.

\begin{figure}[ht!]
\centering
\includegraphics[scale=0.6,clip=true]{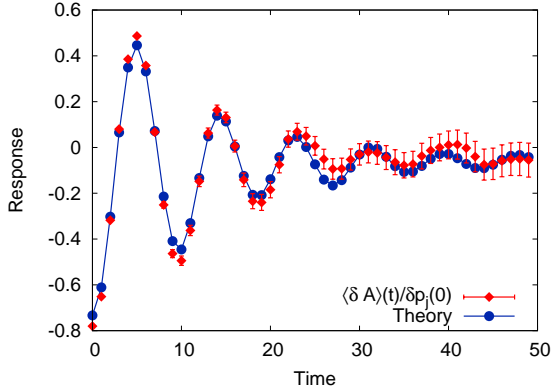}
\caption{Check of the FDR by a direct computation of the mean response
  and comparison with the theory in the model~(\ref{longrange}) for the variable $A=\sin p_j$, for
  $E/N=1.9$. Other parameters are $J=-0.5,\, K=-1.4, \, N=250, \,
  M=10^5$ and $\delta p_i(0)=0.01$~\cite{N2}. Reproduced figure with permission from F. Miceli, M. Baldovin, and A. Vulpiani. Phys. Rev.
E, 99:042152, 2019, Copyright (2019) by the American Physical Society.}
\label{figFDR}
\end{figure}

The generalized FDR derived in Sec.~\ref{GFDRs} holds even for systems
with negative temperature. Since the, somehow, peculiar features of
such systems, the validity of the FDR can appear to be not obvious. As
an example, in Fig.~\ref{figFDR} from~\cite{N2}, we show the
comparison between the response function and the (proper) correlation
functions numerically obtained in a model with long-range interactions
\begin{equation}
  \label{longrange}
H= \sum_{i=1}^N (1- \cos p_i) +
N\Big( {J \over 2} m^2+ {K \over 4} m^4 + const. \Big),
\end{equation}
where $m$ is the modulus of the vector
$$
{\bf m}= \Big( {1 \over N}\sum_{i=1}^N \cos q_i, 
{1 \over N}\sum_{i=1}^N \sin q_i \Big),
$$
for a value of  $E$ corresponding to a negative temperature.

\subsubsection{Gaussian distribution}

When the stationary state is described by a multivariate Gaussian distribution, 
\begin{equation}
\ln \rho({\bf x})= -{1 \over 2} \sum_{i,j}\alpha_{ij}x_i x_j + const.
\end{equation}
with $\{ \alpha_{ij} \}$ a positive matrix, 
the elements of the linear response matrix can be written in terms
of the usual  correlation functions:
\begin{equation}
\label{3.14}
R_{i,j} (t) =\sum_k \alpha_{j,k}
{\Bigl \langle x_i(t) x_k(0) \Bigr \rangle }  \; .
\end{equation}
The above result is nothing but the Onsager regression originally
obtained for linear Langevin equations.

It is interesting to note that there are important nonlinear systems
with a Gaussian invariant distribution, e.g. the inviscid
hydrodynamics~\cite{Kr59,Kr00}, where the Liouville theorem holds, and
a quadratic invariant exists. Therefore one has a quite simple
relation between the responses and the correlations; in spite of this
fact, the dynamics is not linear and the behavior of the correlation
functions is not trivial at all.

\subsubsection{Non Hamiltonian systems}

When the form of $\rho({\bf x})$
is not known, as usually in non-Hamiltonian systems, the relation (\ref{3.9}) does not give detailed
quantitative information.  However, it represents
a connection between the mean response function $R_{i,j}$ and a
suitable correlation function, computed in the non perturbed systems:
\begin{equation}
\label{3.13}
\Bigl \langle x_i(t)f_j({\bf x}(0)) \Bigr \rangle \; , 
\quad \textrm{with} \quad 
f_j({\bf x})=- \frac{\partial \ln \rho}{\partial x_j} \,\, ,
\end{equation}
where, in the general case, the function $f_j$ is unknown.

Let us stress that, in spite of the technical difficulty for the
determination of the function $f_j$, which depends on the invariant
measure, a FDR always holds in mixing systems whose invariant measure
is ``smooth enough''.  In particular, we note that the nature of the
statistical steady state (either equilibrium, or non-equilibrium) has
no role at all for the validity of the FDR.

\subsubsection{Marginal distribution}

Let us stress that the knowledge of a marginal distribution
\begin{equation}
p_i(x_i)= \int \rho(x_1, x_2, ....)\prod_{j \neq i}dx_j
\label{projection} 
\end{equation} 
does not allow  for the computation of the auto-response:
\begin{equation}
R_{i,i}(t)\neq
- \Biggl \langle x_i(t) \left.
 \frac{\partial \ln p_i(x_i)} {\partial x_i} \right|_{t=0}
\Biggr  \rangle \,\,,  \label{falseresponse}
\end{equation}
even in the case of Gaussian variables. As one can easily understand
from Eq.~(\ref{3.14}), $R_{i,i}(t)$ in general, even in the Gaussian
case, is not proportional to $\langle x_i(t)x_i(0) \rangle $.  This
observation is very relevant for what follows.  Consider for instance
that the description of our system has been restricted to a single
\emph{slow} degree of freedom, such as the coordinate of a colloidal
particle following a Langevin equation.  The above discussion has
indeed made clear that other fluctuating variables can be coupled to
the one we are interested in, and therefore it is not correct to
project them out by the marginalization in Eq.~(\ref{projection}).
Conversely, a stationary probability distribution with new variables
coupled to the colloidal particle position must be taken into account.
This point will be discussed in more detail in Sec.~\ref{marginal}.

\subsubsection{Chaotic dissipative systems}

At this point one could object that in a chaotic deterministic
dissipative system the above machinery cannot be applied, because the
invariant measure is not smooth at all.  Typically the invariant
measure of a chaotic attractor has a multifractal structure and its
Renyi dimensions $d_q$ are not constant.  In chaotic dissipative
systems the invariant measure is singular. However the previous
derivation of the FDR relation is still valid if one considers
perturbations along the expanding directions.  Due to the singular
nature of the invariant probability distribution, a general response
function has two contributions, parallel and perpendicular to the
attractor, corresponding respectively to the expanding (unstable) and
the contracting (stable) directions of the dynamics~\cite{N3}.  Each
perturbation can be written as the sum of two contributions
$$
\delta F(t)= \delta F_{\parallel}(t)+ \delta F_{\perp}(t),
$$
and
the effect of such a perturbation on the response on an observable $A$ 
attains the form
\begin{eqnarray}
\overline{ \delta A (t)}&=&
\int_0^t R_{\parallel}^{(A)}(t-t') \delta F_{\parallel}(t') dt' \nonumber \\
&+&
\int_0^t R_{\perp}^{(A)}(t-t') \delta F_{\perp}(t') dt' \, .
\end{eqnarray}
It is easy to realise that only for the part $R_{\parallel}^{(A)}$
one can have a FDR, i.e. it can be expressed in terms of a correlation
function computed in the unperturbed dynamics on the attractor.  On
the contrary for the second contribution (from the contracting
directions), the response can be obtained only
numerically~\cite{CS07}.

As a matter of fact, there are at least two good reasons which allow
us to hope that the singular structure of the invariant measure, at
practical level, can be not very relevant.  First, we note that in
systems with many degrees of freedom, for a non pathological
observable the contribution of $R_{\perp}^{(A)}$ should be negligible
(we will consider again in the following this point)~\cite{N4}.  In addition, a
small amount of noise, that is always present in a physical system,
smooths the $\rho({\bf x})$ and the FDR can be derived.  We recall
that this ``beneficial'' noise has the important role of selecting the
natural measure, and, in the numerical experiments, it is provided by
the round-off errors of the computer. We stress that the assumption on
the smoothness of the invariant measure allows one to avoid subtle
technical difficulties.

\subsection{Other forms of non-equilibrium FDR}

In Sec.~\ref{GFDRs} we have derived a FDR which involves the invariant
measure.  Here we consider a somehow complementary approach, in the
sense that one expresses the response functions in terms of
correlations with a quantity that involves the transition rates of the
model. The common features, namely the appearance of extra-terms
related to the coupling with ``hidden'' variables, will be discussed
in the next subsection.

We consider a system with dynamics described by a Markov process with
transition rates $W({\bf x'}\to{\bf x})$ from state ${\bf x'}$ to
state ${\bf x}$, with normalization
\begin{equation}
\sum_{{\bf x}} W({\bf x'}\to {\bf x})=0.
\label{2.0}
\end{equation}
Assuming a perturbation in the form of a time-dependent external field
$h(s)$, which couples to the potential $V({\bf x})$ and changes the
energy of the system from $H({\bf x})$ to $H({\bf
  x})-h(s)V({\bf x})$, the linear response function of the observable
$A$ is
\begin{equation}
R(t,s)= \left .\frac{\delta
  \langle A(t)\rangle_h}{\delta h(s)}\right|_{h=0},
\label{2.4}
\end{equation}
where $\langle\ldots\rangle_h$ denotes an average on the perturbed
dynamics. The perturbed transition rates $W^h({\bf x}|{\bf x'})$, to
linear order in $h$, take the form
\begin{eqnarray}
W^h({\bf x'}\to {\bf x})&=&W({\bf x'}\to {\bf x}) \times \nonumber \\
&&\left\{1-\frac{\beta h}{2}\left[V({\bf x})-V({\bf x'})\right]+M({\bf x},{\bf x'})\right\},
\label{2.3}
\end{eqnarray}
where $\beta$ is the inverse temperature (with $k_B=1$) and $M({\bf
  x},{\bf x'})$ is an arbitrary symmetric function. The diagonal
elements are obtained from the normalization
condition~(\ref{2.0}). This structure derives from the local detailed
balance principle~\cite{CM99}. Note, however, that there is not a
univocal prescription for the choice of $W^h$ through the function
$M$~\cite{LCZ05,CLSZ10}. For a general discussion of different
symmetric factors in the transition rates, see for
instance~\cite{Basu_2015}, or~\cite{PhysRevE.93.032128,baiesinew} in
the context of lattice gas models. For simplicity, here we take $M=0$.
Then, the response function can be written as
\begin{equation}
R(t,s)= \frac{\beta}{2}\left[\frac{\partial \langle A(t)
    V(s)\rangle}{\partial s} -\langle A(t)B(s)\rangle \right],
\label{2.9}
\end{equation}
where
\begin{equation}
B(s) \equiv B[{\bf x}(s)]=\sum_{{\bf x''}}\{V({\bf x''})-V[{\bf x}(s)]\}W[{\bf x}(s)\to {\bf x''}]
\label{2.10}
\end{equation}
is an observable quantity, namely depends only on the state of the
system at a given time.

The two formulae, Eqs.~(\ref{3.9}) and ~(\ref{2.9}), show that, in
general, non-equilibrium is not a limit, in the sense that the
response function can still be expressed in terms of unperturbed
correlators. Similar forms of FDR have been derived
in~\cite{BMW09,maes1,maes2,maes2,ss06,Seifert,Verley_2011,PhysRevLett.103.090601},
and also experimentally verified~\cite{GPCCG09,GPCM11}. In particular,
Eq.~(\ref{2.9}) extends to discrete systems the relation derived for
overdamped Langevin equation with continuous variables~\cite{CKP94}.
Let us also mention the rigorous derivation of a similar FDR in the
context of exclusion processes reported in~\cite{olla} and the recent
results for the response to temperature
perturbations~\cite{falasco,falasco2}.

A simple illustration of Eq.~(\ref{2.9}) is provided by the case of
the Langevin dynamics of a particle diffusing in a potential $U(x)$,
\begin{equation} 
\dot{x}(t) = - \frac{\partial U(x)}{\partial x}+ \sqrt{2 T } \zeta(t), 
\label{2.10bis}
\end{equation}
with $\zeta(t)$ a white noise with zero mean and unit variance. The
response formula, with respect to a perturbing force $F$,
reads as:
\begin{equation}
\frac{\delta \langle x(t) \rangle_F}{\delta F(s)} = 
\frac{\beta}{2} \left[ \frac{\partial \langle x(t) x(s) \rangle}{\partial s} - \langle x(t) B[x(s)] \rangle \right], 
\label{2.12}
\end{equation}
with $B[x(s)] = -\partial U/\partial x|_{x(s)}$.  At equilibrium it
can be easily proved that $\langle x(t) B[x(s)] \rangle = - \partial
\langle x(t) x(s) \rangle / \partial s $, recovering the standard FDR
formula. Differently, out of equilibrium the contribution coming from
the local field $B[x(s)]$ must be explicitly taken into account.

Finally, we note that, in this theoretical approach, non-linear FDRs
can be derived, relating non-linear response functions with high-order
correlation
functions~\cite{bouchaud2005nonlinear,LCSZ08a,LCSZ08b}. These
nonlinear responses find important applications in the context of
disordered and glassy systems~\cite{crauste2010evidence}. An important
point to stress is that, for non-linear FDRs, even at equilibrium the
model dependent quantity $B$ defined in Eq.~(\ref{2.10}) is involved.
These results point out the central role played by kinetic factors in
characterizing the non-equilibrium dynamics~\cite{Basu}.

\subsection{Usually marginal PDF is not enough: the role of coupling}
\label{marginal}

The main message from the previous section is that out of equilibrium the
response function can still be expressed in terms of unperturbed
correlators but these correlators involve particular quantities that
do not appear at equilibrium.  These quantities indeed characterize
the non-equilibrium dynamics.

There is still a deep debate on the general physical meaning of such
terms: some ones have pointed out the role played by different entropy
productions~\cite{seifert05,ss06}, some others have pinpointed the
necessity to introduce new perspectives in order to characterize
non-equilibrium dynamics~\cite{maesbook}.

What is clear is that the extra-terms unveil the presence of relevant
couplings that arise in non-equilibrium systems. In particular, even
if one is interested in the response of a specific variable, the
knowledge of its statistical properties, namely the marginalized
distribution function, is in general not enough. Other degrees of
freedom can be coupled to the observable under study, making the
prediction of its response more involved.

\begin{figure}[ht!]
\centering
\includegraphics[scale=0.3,clip=true]{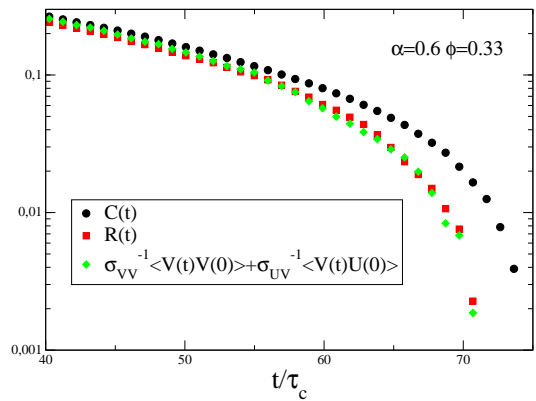}
\caption{Velocity correlation $C(t)=\langle V(t)V(0)\rangle/\langle
  V(0)V(0)\rangle$ (black circles), response function (red squares),
  and generalized FDR Eq.~(\ref{gfdrtracer}) (green diamonds) computed
  in molecular dynamics numerical simulations of a massive tracer in a
  granular gas with packing fraction $\phi=0.6$, coefficient of
  restitution $\alpha=0.6$ and collision time $\tau_c$,
  see~\cite{SVGP10} for details.}
\label{fdrtracer}
\end{figure}

An illustrative example is provided by the motion of an intruder in a
granular gas~\cite{SVCP10,SVGP10,GPSV14}. A granular fluid is made of
macroscopic particles subject to external forcing, and therefore is
characterized by dissipative interactions (inelastic collisions) and
non-equilibrium dynamics.  In order to describe the velocity
autocorrelation of the intruder and its linear response, one can
introduce a two-variable model (in one dimension, for simplicity)
\begin{subequations} \label{grintr}
\begin{align} 
M \dot{V}(t)= - \Gamma [V(t)-U(t)] + \sqrt{2 \Gamma T_{tr}} \mathcal{E}_v(t)\\
M' \dot{U}(t) = -\Gamma' U(t) - \Gamma V(t) + \sqrt{2 \Gamma' T_b} \mathcal{E}_U(t),
\end{align}
\end{subequations}
where $V$ is the velocity of the intruder with mass $M$, $U$ describes
a local velocity field (a local average of the velocities of the
particles surrounding the intruder) whose dynamics is coupled with
that of the tracer, $\Gamma$ is a viscosity, $\Gamma'$ and $M'$ are
effective parameters to be determined. $T_{tr}$ is the intruder
kinetic temperature, while $T_b$ is the value of the kinetic
temperature of the granular fluid, playing the role of a
non-equilibrium bath. $\mathcal{E}_v$ and $\mathcal{E}_U$ are
delta-correlated noises with zero mean and unitary variance. The
dilute limit can be obtained with $\Gamma' \sim M' \to \infty$, which
implies small $U$.

Eqs.~\eqref{grintr} represent a linear model for which analytical
results can be obtained and can describe real systems for not too high
density.  In particular, in the elastic limit ($T_{tr} = T_b$), the
coupling with $U$ can still be important, but the equilibrium FDR is
recovered.  Out of equilibrium, one can apply the formula~(\ref{3.9})
to express the response in terms of correlation functions. Since the
system is linear, the stationary distribution is a bivariate Gaussian,
and from Eq.~(\ref{3.9}) directly follows
\begin{equation}
 R_{VV}(t)= \frac{\overline{\delta V(t)}}{\delta V(0)}=\sigma_{VV}^{-1}\langle V(t)V(0)\rangle + \sigma_{UV}^{-1}\langle V(t)U(0)\rangle,
  \label{gfdrtracer}
\end{equation}
where $\sigma_{VV}^{-1}$ and $\sigma_{UV}^{-1}$ are the elements of
the inverse covariance matrix and can be expressed in terms of the
model parameters, see~\cite{CPV12} for further details. In
Fig.~\ref{fdrtracer} we check the validity of such an approach. The
main message we want to stress here is that a central role is played
by the correlations between the variable $V$ and the local velocity
field $U$. At variance with equilibrium cases, in general the
knowledge of the statistical properties of $V$ alone, e.g. the measure
of its marginalized PDF, is not enough to reconstruct the response to
an external perturbation.

\section{Toward realistic systems}
\label{real}

In this Section we illustrate the use of FDRs introduced above, in
different non-equilibrium and non-standard systems, ranging from
athermal and active matter to systems with anomalous diffusion and
multiple time-scale structure.

\subsection{FDR and the effective temperature}

One of the applications of the FDR in non-equilibrium systems deals
with the interesting concept of \emph{effective}
temperature~\cite{CKP97}.  Indeed, within the context of
non-equilibrium phenomena, the first attempts to formulate a general
theory start from extending concepts well defined in the consolidate
equilibrium theory. Here, for instance, one assumes that, after a
sufficient long time $t_{eq}$, an isolated and finite system reaches
an equilibrium state that can be characterized by a small number of
parameters, the state variables, such as temperature and pressure.
Thus, an interesting question concerns the possibility that also for
non-equilibrium phenomena, a characterization in terms of a few
variables is still feasible, at least in some particular regimes.

The effective temperature can be introduced via the linear FDR, as the
ratio between response function and correlation function~\cite{CKP97}.
The study of the behavior of such a quantity can be interesting in
itself, but its real meaning as relevant parameter characterizing some
features of the system strongly depends on the considered models.  The
shape of the fluctuation-dissipation ratio can be useful to grasp
information on the presence of different relevant time scales in the
system. However, the observed different time scales are not
  necessarily related to an underlying complex
dynamics~\cite{VBPV09}.  The general issue of effective temperature
has been the subject of recent reviews~\cite{cuglirev,PSV17} and is an
open line of intense research, with applications for instance in
granular~\cite{DMGBLN03,BBDLMP05,keys2007measurement} and active
matter~\cite{PhysRevE.77.051111,Wang15184,PhysRevE.90.012111,Levis_2015,Han7513,PhysRevE.97.032125,dieterich2015single,preisler2016configurational,PhysRevLett.118.015702,workamp2018symmetry,seifert2019stochastic,cugliandolo2019effective,golestanian,klongvessa}
(see also Section~\ref{actives} below).  Here, we focus on a few
examples in the context of athermal systems.

First, we mention some cases where the idea of effective temperature
has been successfully applied. It has been shown that, in a model of a
sheared, zero-temperature foam, different definitions of temperature,
that for equilibrium thermal systems would be equivalent, take on the
same value and show the same behavior as a function of the
shear-rate~\cite{durian1}. This observation suggests that in this
situation the concept of effective temperature is robust and its
introduction can be useful to build a statistical mechanics
description out of equilibrium. Similar conclusions followed for other
athermal systems, such as a sphere placed in an upward flow of
gas~\cite{durian2} and a two-dimensional air-driven granular
medium~\cite{durian3}.

\begin{figure}[ht!]
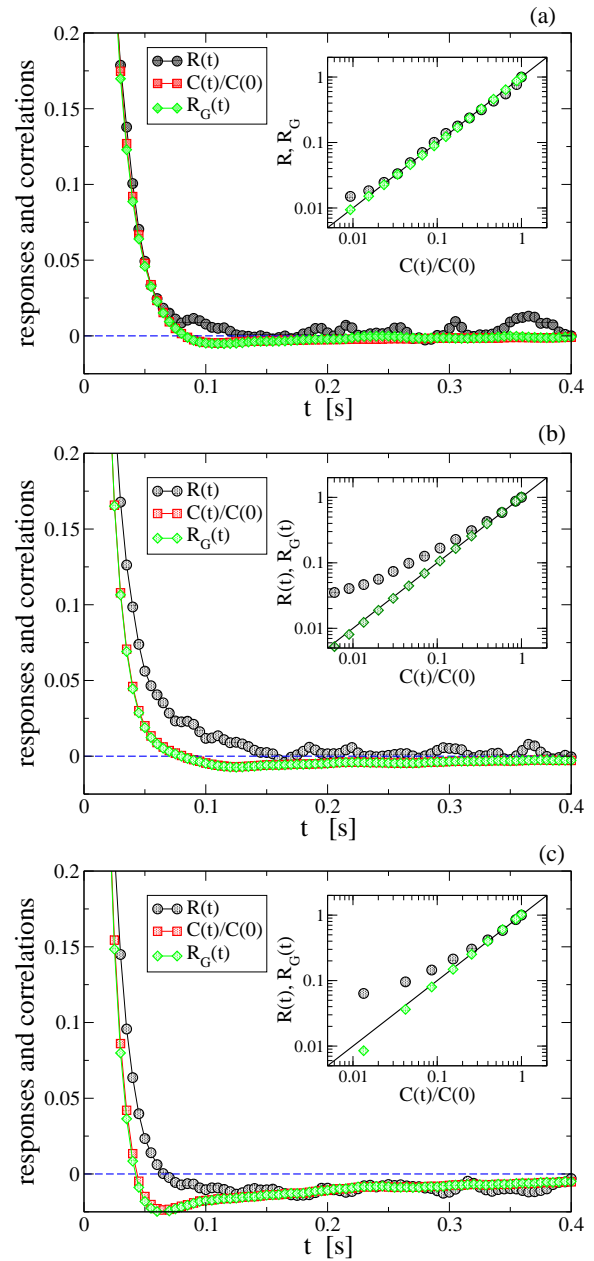

\centering
\includegraphics[scale=0.3,clip=true]{new_fig_dil.eps}
\includegraphics[scale=0.3,clip=true]{new_fig_med.eps}
\includegraphics[scale=0.3,clip=true]{new_fig_den.eps}
\caption{Response function $R(t)$, velocity autocorrelation $C(t)$ and
  predicted response with a factorization approximation $R_G(t)$,
  measured for a blade suspended in a strongly vibrated granular gas~\cite{GPSV14}
  for three different packing fractions (0.05 (a), 0.1 (b), and 0.15 (c)),
  reproduced with permission from A. Gnoli, A. Puglisi, A. Sarracino, and A. Vulpiani. Plos One, 9:e93720, (2014). 2014 Creative Commons Attribution (CC BY) license}
\label{fdrexp}
\end{figure}

In a different experimental setup the response and autocorrelation
functions of a blade suspended in a strongly vibrated granular system
were measured~\cite{GPSV14}. In this case, it has been shown that only
in the dilute regime an effective temperature can be properly defined,
being the response and autocorrelation proportional to each other. At
higher densities, on the contrary, response and correlations show more
complex behaviors, see Fig.~\ref{fdrexp}, and the coupling with other
degrees of freedom in the system starts to play a central role.  In
particular, inspired by the model described in Sec.~\ref{marginal}, it
was suggested that the relevant coupling quantities can be defined in
terms of a local velocity field.  Experimental measures confirmed such
an explanation, even if a quantitative description of the system
correlations with the simple two-variable model turned out to be not
accurate.

\subsection{FDR and anomalous diffusion}
\label{comb}

Another non-standard class of systems where the FDR has
been investigated is represented by systems showing anomalous
transport properties. These systems are characterized by a non-linear behavior
of the mean square displacement~\cite{BG90,GSGWS96,CMMV99,MK00,BC05,BO02}, i.e.
\begin{equation}
\label{4}
\langle x^2(t)\rangle \sim t^{2 \nu} \,\,\, \mbox{with} \,\,\,  \nu \ne
1/2.
\end{equation}
Formally this corresponds to have a diffusion coefficient $D=\infty$
if $\nu > 1/2$ (superdiffusion) and $D=0$ if $\nu < 1/2$
(subdiffusion). Note that in general the same model can show different
behaviors depending on the considered time scale.  It is interesting
to wonder whether the FDR in the form of the Einstein relation is
still valid, namely whether the quantity $\langle x^2(t) \rangle$ is
still proportional to $\overline{\delta x}(t)$ at any time:
\begin{equation}
\label{3}
{\langle x^2(t) \rangle  \over \overline{\delta x}(t)}=\frac{2}{\beta F},
\end{equation}
  where
  \begin{equation}
\label{2}
\overline{\delta x}(t)=
\langle x(t) \rangle_F - \langle x(t) \rangle
 \simeq \mu F t \,\, ,
\end{equation}
with $\langle\ldots\rangle_F$ denoting the average on the system
perturbed by a force $F$, and $\mu$ the mobility.

Quite remarkably, it has been shown that the Einstein relation is
robust and holds even in models showing anomalous behaviors. This has
been explicitly proved in systems described by a
fractional-Fokker-Planck equation~\cite{MBK99,BF98,CK09}. In addition
there is clear analytical~\cite{LATBL10} and numerical~\cite{VPV08}
evidences that~(\ref{3}) is valid for the elastic single file model,
i.e. a one-dimensional gas of elastic particles on a ring, which
exhibits subdiffusive behavior due to the confinement, $\langle x^2
\rangle \sim \sqrt{t}$~\cite{HKK96}.

\begin{figure}[t!]
\centering
\includegraphics[scale=0.3,clip=true]{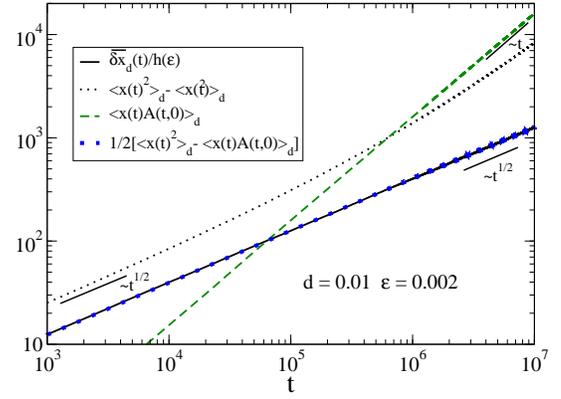}
\caption{Response function (black line) and second cumulant (black dotted line) measured in the
  comb model~\cite{comb}.  The term including the correlation with the quantity
  $B$ (green dotted line) is necessary to recover the response
  function (see blue dotted line), in agreement with the
  FDR~(\ref{2.9}). }
\label{fdrcomb}
\end{figure}

An important point to stress is that the validity of a FDR in the
standard Einstein form depends on the equilibrium properties of the
systems, rather than on its anomalous dynamics. Indeed, if
non-equilibrium conditions are introduced in the system, a generalized
FDR has to be considered, including the extra-terms previously
discussed. This has been explicitly shown, for instance, for a
particle diffusing on a comb lattice~\cite{comb,Forte_2013}, driven by
an external force.  In particular, denoting by $x\in(-\infty,\infty)$
the position of the particle along the backbone of the comb and by
$y\in[-L,L]$ the coordinate along a tooth, transition rates from
$(x,y)$ to $(x',y')$ are
\begin{eqnarray}
W^d[(x,0)\rightarrow (x\pm 1,0)]&=&1/4\pm d \nonumber \\
W^d[(x,0)\rightarrow (x,\pm 1)]&=&1/4 \nonumber \\
W^d[(x,y)\rightarrow (x,y\pm 1)]&=&1/2~~~ \textrm{for}~y\ne 0,\pm L,
\label{ww}
\end{eqnarray}
where $d$ is the drift in the $x$ direction.  Applying the generalized
FDR Eq.~(\ref{2.9}) for the response to a perturbation $\varepsilon$
one has~\cite{comb}
\begin{eqnarray}
\hspace*{-2cm}\frac{\overline{\delta\mathcal{O}}_d}{h(\varepsilon)}&=&\frac{\langle \mathcal{O}(t)\rangle_{d+\varepsilon}- \langle
\mathcal{O}(t)\rangle_d}{h(\varepsilon)} \nonumber \\
&=&\frac{1}{2}\left[\langle\mathcal{O}(t)x(t)\rangle_d-\langle\mathcal{O}(t)x(0)\rangle_d
-\langle\mathcal{O}(t)A(t,0)\rangle_d\right],
\label{FDR}
\end{eqnarray}
where $\mathcal{O}$ is a generic observable, and $A(t,0)=\sum_{t'=0}^t
B(t')$, with $B$ following from Eq.~(\ref{2.10})
\begin{equation}
B[(x,y)]=\sum_{(x',y')}(x'-x)W^d[(x,y)\rightarrow (x',y')]=2d\delta_{y,0}.
\end{equation}
The sum on $B$ has an intuitive meaning: it counts the time spent by
the particle on the backbone.  In Fig.~\ref{fdrcomb} it is shown the
validity of this approach.

Let us note that in these cases the FDR holds if the displacement is
compared to the linear response but it does not if the
self-correlation of the particles position is used, due to the lack of
a confining potential.

Another example is the inelastic single-file model~\cite{VPV08}, where
one introduces dissipative interactions among the particles diffusing
on a ring. Again, due to the non-equilibrium conditions, strong
correlations among particles are present and the factorization of the
invariant measure fails.  For small dissipation (namely, small
inelasticity, small packing fraction and/or fast thermostats) the
Einstein relation is recovered, because of the weak lack of
factorization.  In general, the FDR involves extra-terms that take
into account correlations with other degrees of freedom in the system.

Generalized FDR have also been applied in the case of superdiffusion
for instance in models of L\'evy walks and L\'evy
flights~\cite{Gradenigo_2012,PhysRevE.87.030104,godec2013linear,kusmierz2018thermodynamics}
or generalized Langevin
equations~\cite{costa2003fluctuation,dieterich2015fluctuation}, while
the linear response of a particle showing anomalous diffusion in an
aging medium has been studied in~\cite{pottier,pottier1}.  A very
interesting open problem is the validity of the Einstein relation in
systems of interacting particles on comb structures, where the
diffusion of a tracer can be studied with analytical approaches and
shows different non-trivial anomalous
behaviors~\cite{PhysRevLett.115.220601}. Finally, let us mention that
for a tracer advected by a steady laminar flow and subject to an
external force, showing non-trivial anomalous behaviors, such as
negative differential and absolute mobility, the validity of a
generalized FDR~\cite{maes2} has been recently discussed
in~\cite{PhysRevLett.117.174501}.

\subsection{FDR and multiscale systems}

Let us now discuss some aspects of the FDR in systems with non-trivial
temporal structures, for instance with many degrees of freedom whose
characteristic times are very different.  The relation~(\ref{3.9})
suggests that, in general, the choice of the observable $A({\bf x})$,
and the size of the perturbation $\Delta {\bf x}(0)$, can correspond
to different relaxation behaviours of $\overline{\delta A(t)}$.

In systems with a unique characteristic time, e.g. the celebrated
$3-d$ Lorenz system or $1d$ Langevin equation, one expects that there
are not significant differences at varying the observable and the
size of the perturbation, and numerical computations confirm the
intuition.  Less trivial is the case of high dimensional systems with
many different characteristic times, where one observes a more
interesting scenario.  At varying the observables one has different
correlation functions, whose relaxation times can be very different;
in an analogous way different response functions can be characterized
by very different temporal behaviours.

The fact that finite perturbations can relax with characteristic times
which can depend very much on the size of the initial perturbation, is
rather relevant in geophysical context, e.g. in the study of climate
dynamics, where many degrees of freedom are involved with characteristic
times which vary from seconds (3D turbulence) to weeks (geostrophic
turbulence) and thousand years (oceanic currents and ice shields
dynamics).  Numerical computations on simplified models, e.g. the so
called shell models, show the following scenario: the relaxation time
of the finite perturbations increases with the size at the initial
time~\cite{BLMV03}.

As already discussed, the relations between response and correlation
are not trivial at all: this because there appears the invariant
probability density which is not known, a part very few special cases
(e.g. Hamiltonian systems, inviscid hydrodynamics and some Langevin
equations).  In dissipative systems, as mentioned before, we
have an additional technical difficulty due to the singular structure
of $\rho$.  Nevertheless, we have the positive fact that the
generalized FDR~(\ref{3.9}) indicates the existence of a relation
between the response and some correlation whose precise functional
shape is not known.  It seems natural to hope that the simple
correlations, i.e. $\langle x_i(t)x_i(0) \rangle$ is enough to catch
at least the qualitative behaviour of the response function
$R_{i,i}(t)$.

In order to illustrate this issue, let us briefly discuss a system introduced by Lorenz as a simplified
model for the atmospheric circulation~\cite{N5}:
$$
{d x_k \over dt}=-
x_{k-1}(x_{k-2}-x_{k+1}) -\nu x_k + F + c_1\sum_{j=1}^M y_{k,j}
$$
$$
{d y_{k,j} \over dt}=-
c b y_{k, j+1}(y_{k,j+2}-y_{k,j-1}) -c \nu y_{k,j} + c_1 x_k
$$
where the set $\{ x_k \}$ with $k=1, ... , N$ and
$\{ y_{k,j} \}$ with $j=1, ... , M$ are the slow-large scale
and the fast small-scale variables respectively; the above system
is often called Lorenz-96 model.
Roughly speaking, the $\{ x_n \}$'s represent the synoptic scales, while the
$\{ y_n \}$'s represent the  convective scales.
\begin{figure}[t!]
\centering
\includegraphics[scale=0.3,clip=true,angle=-90]{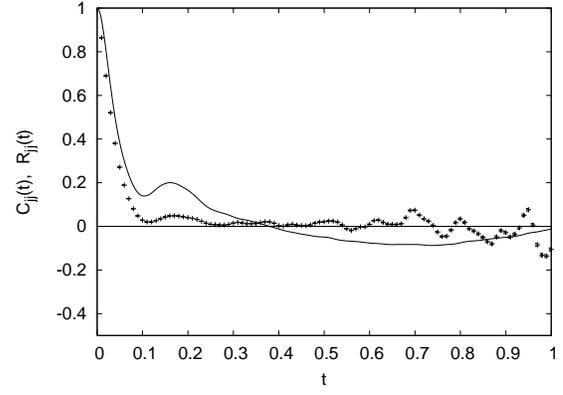}
\caption{Autocorrelation $C_{jj}(t)$ (full line) and
self-response $R_{jj}(t)$ of the fast variable
$y_{k,j}$ for $j=3\, k=3$ for the Lorenz-96 model~\cite{N5};
the parameters are $ F=10,\, \nu=1, \, c=10, \, c_1=1, \,
N=36$ and $M=10$. Reproduced with permission from G. Lacorata and A. Vulpiani. Nonlinear Proc. in Geophys.,
14:681, (2007). Copyright 2007 European Geosciences Union}
\label{figFDR1}
\end{figure}
\begin{figure}[t!]
\centering
\includegraphics[scale=0.3,clip=true]{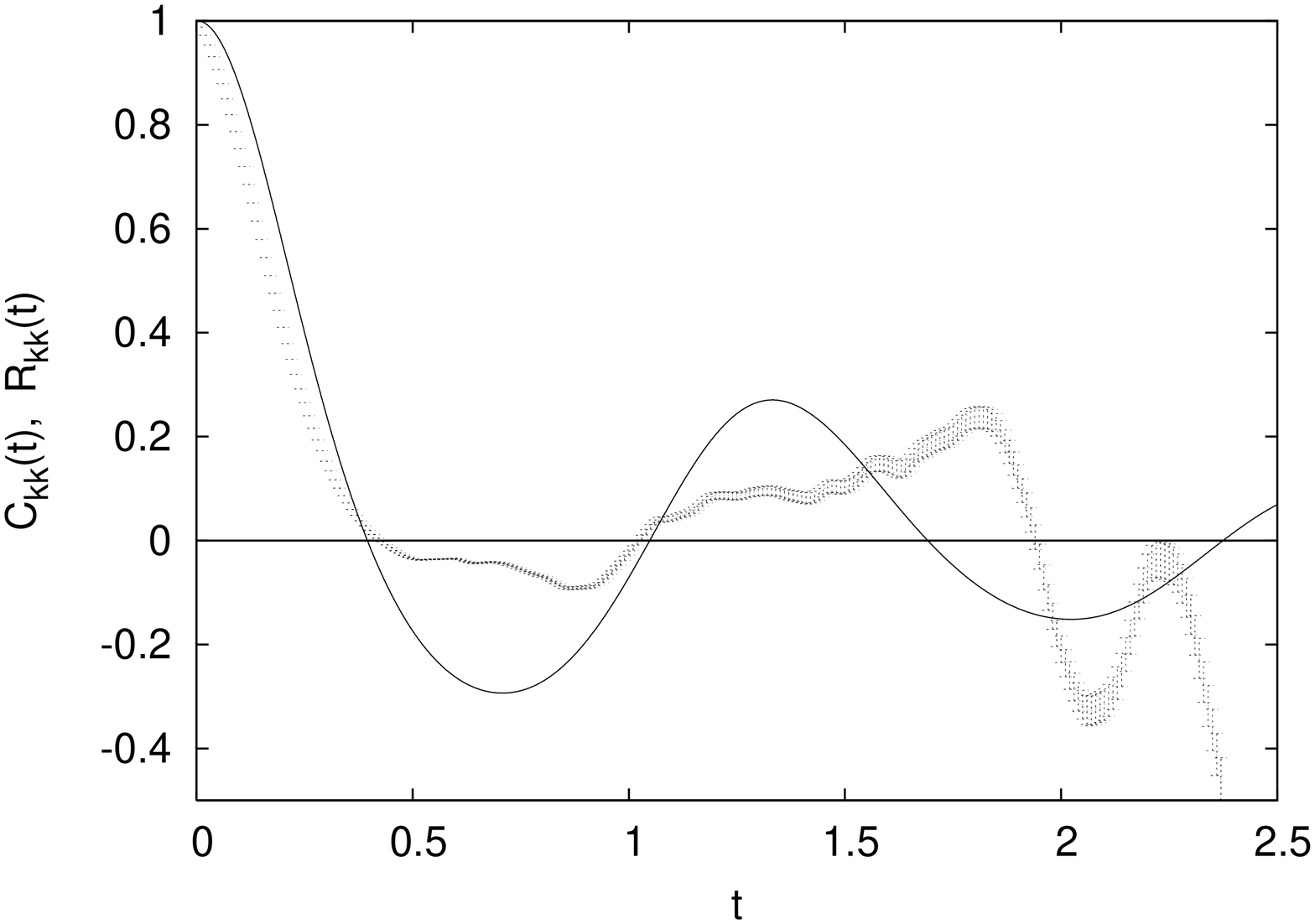}
\caption{Autocorrelation $C_{kk}(t)$ (full line) and
self-response $R_{kk}(t)$ of the slow variable
$x_{k}$ for $k=3$ for the Lorenz-96 model~\cite{N5};
the parameters are $ F=10,\, \nu=1, \, c=10, \, c_1=1, \,
N=36$ and $M=10$. Reproduced with permission from G. Lacorata and A. Vulpiani. Fluctuation-response relation and modeling in systems with fast and slow dynamics. Nonlinear Proc. in Geophys.,
14:681, (2007). Copyright 2007 European Geosciences Union}
\label{figFDR2}
\end{figure}
Of course it is impossible to write down the invariant measure $\rho$
and therefore to find the proper correlation function for a given
response. Nevertheless, as shown in Figs.~\ref{figFDR1}
and~\ref{figFDR2}, it is well evident how, even in the absence of a
precise quantitative agreement, one has a certain similarity between
the autocorrelation and self-response function.  The correlations of
the slow (fast) variables have a clear qualitative resemblance with
the response of the slow (fast) variables themselves.  In particular,
the relaxation times of the response of fast (slow) variables are of
the same order of magnitude of the corresponding correlation
functions~\cite{N5}.

In some applications it is rather common to wonder about the response
of a global variable which depends on many variables: for instance in
the study of climatic dynamics it is rather natural to study the
global temperature.  The intuition suggests that in the presence of
many degrees of freedom, even for dissipative systems, if we are
interested in the response of some global variable, one can hope to
invoke the help of some statistical regularization.
Numerical computations indicate that such an intuition is correct.

Indeed,  in some cases 
the above intuition is confirmed by
a rigorous analytical treatment, as described in 
a recent paper by Wormell and Gottwald~\cite{Wormell}.
In this work it is studied the macroscopic variable $Q$ ruled by the discrete time map
$$
Q_{t+1}=A Q_t(1- Q_t),
$$
with
$$
A=A_0+{A_1 \over M^{\gamma}}\sum_{j=1}^Mx_t^{(j)},
$$
where the independent variables $\{ x^{(j)} \}$ evolve according to a
 deterministic law
$$
x_{t+1}^{(j)}=g_{a_j}(x_t^{(j)}),
$$
e.g. $g_a(x)=x^2- (a x(1-x))^2$, the $\{ a_j \}$ are sampled by a
smooth PDF and $\gamma \ge 1/2$.  The stationary joint PDF of $(Q, \{
x^{(j)} \})$ does not depend in a continuous way on the $\{ a_j \}$
and therefore a FDR cannot hold for the variables $\{ x^{(j)} \}$.  In
spite of this, one has that the marginal PDF of the macroscopic
variable $Q$ varies in a continuous (and differentiable) way with the
$\{ a_j \}$: this allows to show a FDR for the variable $Q$.

\subsection{FDR and active matter}
\label{actives}

In recent years, the field of active matter has drawn the attention of
physicist on the study of a new form of intrinsically non-equilibrium
systems, made by elementary constituents that continuously convert
energy into motion~\cite{DEMAGISTRIS201565}.  This class of systems
shows phenomena similar to those characterizing granular matter, such
as clustering, segregation, non-equilibrium phase transitions,
flocking, collective motion and so on, and many models have been
proposed to describe such a huge variey of behaviors.  Recently, the
generalized FDRs, in different forms, have been applied to these
models, see for
instance~\cite{PhysRevLett.117.038103,PhysRevLett.119.258001,PhysRevE.98.020604,e19070356,PhysRevE.90.052130,maggi,Szamel_2017,chong,burk,pagona}.
Here we mention two specific examples, in order to illustrate some
peculiarities of this context.

As a first example we consider the active Ornstein-Uhlenbeck model,
which describes the persistent motion characterizing a single active
particle, with the introduction of a coloured noise~\cite{maggi1}.
The particle dynamics is modeled as
\begin{eqnarray}
\label{eq:dim_AOUP}
\dot{x}& =& \frac{f(x)}{\gamma}+a, \qquad f(x)=-\frac{d}{dx}\phi(x), \nonumber \\
\dot{a}&=&-\frac{a}{\tau} +\frac{\sqrt{2D}}{\tau}\eta,
\end{eqnarray}
where $x(t)$ is the position of the particle, $\tau$ is the
persistence time, $\gamma$ the drag coefficient, $\phi(x)$ the
potential acting on the system, $a(t)$ is the active bath, and
$\eta(t)$ a delta-correlated white noise, with zero mean and unitary
variance.  The parameter $D$ fixes the amplitude of the active bath
fluctuations
\begin{equation}
\left<a(t)a(t') \right> = \frac{D}{\tau} \exp{\left[-\frac{t-t'}{\tau}\right]}.
\end{equation}
The response to an external perturbation of a one-dimensional system
of non-interacting particles described in this framework has been
recently studied in~\cite{Caprini_2018}. A direct application of
Eq.~(\ref{3.9}) is possible in this case because the stationary
distribution can be computed with some approximations.  In particular,
it has been shown that the non-equilibrium coupling between particle
velocity and position has to be taken into account, playing a central
role when the particle persistence time is large. Moreover, the
analysis showed that although the approximation for the stationary
distribution gives good results for the static properties, the
dynamical behavior can be well described only in the limit of small
persistence time. An FDR for Langevin equations with memory has been
derived in~\cite{safaverdi} from a different approach.

\begin{figure}[t!]
\centering
\includegraphics[scale=0.3,clip=true]{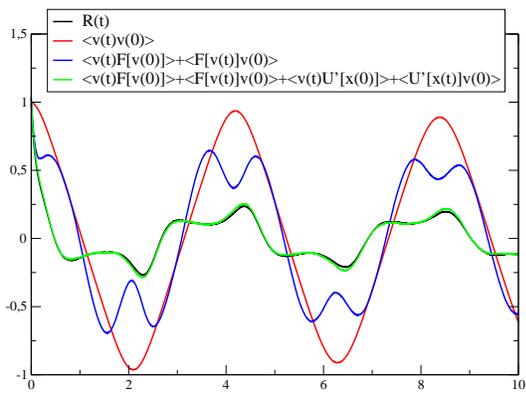}
\caption{FDR Eq.~(\ref{resp.0}) for the model of active
  particles~(\ref{active}) with parameters $\gamma_1=5, \gamma_2=1,
  \gamma T=0.5$ for a potential $U(x)=(1/2)x^2+(1/3)x^3+(1/4)x^4$.}
\label{fdractive}
\end{figure}

A different approach is the active Brownian particle model with energy
depots~\cite{PhysRevLett.80.5044}, derived from the Rayleigh-Helmholtz
treatment of sustained sound waves~\cite{strutt1945theory}.
We consider the equation for the motion of a
particle of mass $m=1$, with position $x$ and velocity $v$, 
in an external potential $U(x)$ 
\begin{eqnarray}\label{kramers}
\dot{x}(t) &=& v(t) \nonumber \\
\dot{v}(t)&=&-F[v(t)]-U'[x(t)] +\eta(t), 
\end{eqnarray}
where $\eta(t)$ is a white noise, with zero mean and $\langle
\eta(t)\eta(t')\rangle=2\gamma T\delta(t-t')$, $\gamma$ and $T$ being
two parameters and $U(x)$ is a potential. The function $F[v]$ is given
by
\begin{equation}\label{active}
F[v(t)]=-\gamma_1 v(t)+\gamma_2 v^3(t),
\end{equation}
with $\gamma_1$ and $\gamma_2$ positive constants.  This means that
motion of the particle is accelerated at small $v$ and is damped at
high $v$, taking into account the internal energy conversion of the
active particles coupled to other energy sources. The stochastic
equations~(\ref{kramers}) are out of equilibrium and a generalized FDR
similar to Eq.~(\ref{2.9}) can be applied~\cite{PhysRevE.88.052124}.
In particular, the response function $R(t,t')$ of the velocity $v(t)$
to a perturbation $h(t')$ applied at a previous time $t'$ reads
\begin{equation}
R(t,t')=\left .\frac{\delta \langle v(t)\rangle}{\delta
  h(t')}\right|_{h=0},
\label{fdr.00}
\end{equation}
and exploiting the relation valid for Gaussian noise~\cite{CKP94}
\begin{equation}
R(t,t')=\frac{1}{2\gamma T}\langle v(t)\eta(t')\rangle,
\label{fdr.0}
\end{equation}
we get the FDR
\begin{eqnarray}
R(t)&=&\frac{1}{2\gamma T}\Big\{\langle v(t)F[v(0)]\rangle 
+\langle F[v(t)]v(0)\rangle  \nonumber \\
&+&\langle v(t)U'[x(0)]\rangle+
\langle U'[x(t)]v(0)\rangle\Big\}.
\label{resp.0}
\end{eqnarray}
%% This expression represents an extension to inertial cases of the
%% result obtained in~\cite{CKP94} for overdamped dynamics.  Note that
%% in the case of a quadratic potential $U(x)=kx^2/2$, Eq.~(\ref{resp.0})
%% can be further simplified, exploiting the relation $\langle
%% x(t)v(0)\rangle=-\langle v(t)x(0)\rangle$, and one gets
%% \begin{equation}
%% R(t)=\frac{1}{2\gamma T}\left\{\langle v(t)F[v(0)]\rangle +\langle
%% F[v(t)]v(0)\rangle\right\}.
%% \end{equation}
%% When an \emph{equilibrium} stationary state is attained, using the
%% Onsager reciprocity relations $\langle v(t)F[v(0)]\rangle= \langle
%% F[v(t)]v(0)\rangle$ and $\langle v(t)U'[x(0)]\rangle=-\langle
%% U'[x(t)]v(0)\rangle$, one finds the EFDR
%% \begin{equation}
%% R_{EFDR}(t)=\frac{1}{\gamma T}\langle v(t)F[v(0)]\rangle.
%% \label{resp2.0}
%% \end{equation}
In Fig.~\ref{fdractive} we show the validity of the above formula,
where, again, the non-equilibrium coupling between velocity and
position has to be taken into account.

\section{Conclusions and perspectives}
\label{conc}

We have reviewed a series of results on FDRs for out-of-equilibrium
systems. The {\em leitmotiv} of our discussion is the importance of
correlations among different variables for non-equilibrium response,
much more relevant than in equilibrium systems: indeed the generalized
FDRs discussed in Section~\ref{GFDR} deviate from the equilibrium
counterpart for the appearance of additional contributions coming from
correlated degrees of freedom. This is the case, for instance, of the
linear response in the general two-variable Langevin model for a
granular tracer: additional contributions to the equilibrium linear
response appear when the main field $V$ is coupled to the auxiliary
field $U$, which only appear out of equilibrium, see
Section~\ref{marginal}, or in the diffusion model on a comb-lattice,
where -- in the presence of a net drift -- the linear response takes a
non-negligible additive contribution, see Section~\ref{comb}. Other
examples have been discussed in Section~\ref{actives}.

Remarkably, in some cases, one may explicitly verify that the coupled
field which contributes to the linear response in non-equilibrium
setups is also involved in the violation of detailed balance: such a
violation is measured by the fluctuating entropy production, whose
connection with non-equilibrium couplings represents another
interesting line of research~\cite{Seifert}.

In conclusion, many results point in the same direction, suggesting a
general framework for linear-response in systems with non-zero entropy
production. Even in out-of-equilibrium dynamics, a clear connection
between response and correlation in the unperturbed system exists. A
further step is looking for more accessible observables for the
prediction of linear response: indeed, both the discussed formula,
Eq.~\eqref{3.9} and Eq.~\eqref{2.9}, require the measurement of
variables which, in general, depend on full phase-space (microscopic)
observables and can be strictly model-dependent. Such a difficulty
also explains why, in a particular class of slowly relaxing systems
with several well separated time-scales, such as spin or structural
glasses in the aging dynamics, i.e. after a sudden quench below some
dynamical transition temperature, approaches involving ``effective
temperatures'' have been used in a more satisfactory
way~\cite{CR03}. Other interpretations of the additional
non-equilibrium contributions to the FDR have been proposed
recently~\cite{maesbook}, but the predictive power of this approach is
not yet fully investigated and represents an interesting line of
ongoing research.

\begin{acknowledgments}

We thank A. Baldassarri, G. Boffetta, F. Corberi, M. Falcioni,
G. Gradenigo, G. Lacorata, E. Lippiello, U. Marini Bettolo Marconi,
A. Puglisi, L. Rondoni, D. Villamaina and M. Zannetti for a long
collaboration on the issues here discussed. AS acknowledges support
from ``Programma VALERE'' from University of Campania
``L. Vanvitelli''.
\end{acknowledgments}

\bibliography{fluct.bib}

\begin{thebibliography}{100}

\bibitem{E05}
A~Einstein.
\newblock On the movement of small particles suspended in a stationary liquid
  demanded by the molecular-kinetic theory of heat.
\newblock {\em Ann. d. Phys.}, 17:549, 1905.

\bibitem{O31}
L~Onsager.
\newblock Reciprocal relations in irreversible processes. {I}.
\newblock {\em Phys. Rev.}, 37:405--426, 1931.

\bibitem{O31b}
L~Onsager.
\newblock Reciprocal relations in irreversible processes. {II}.
\newblock {\em Phys. Rev.}, 38:2265--2279, 1931.

\bibitem{Kr00}
R~H Kraichnan.
\newblock Deviations from fluctuation-relaxation relations.
\newblock {\em Physica {A}}, 279:30, 2000.

\bibitem{Kr59}
R~H Kraichnan.
\newblock Classical fluctuation-relaxation theorem.
\newblock {\em Phys. {R}ev.}, 113:118, 1959.

\bibitem{L75}
C~E Leith.
\newblock Climate response and fluctuation dissipation.
\newblock {\em J. {A}tmos. {S}ci.}, 32:2022, 1975.

\bibitem{B80}
R~E Bell.
\newblock Climate sensitivity from fluctuation dissipation: Some simple model
  tests.
\newblock {\em J. {A}tmos. {S}ci.}, 37:1700, 1980.

\bibitem{NBH93}
G~R North, R~E Bell, and J~W Hardin.
\newblock Fluctuation dissipation in a general circulation model.
\newblock {\em Climate {D}ynamics}, 8:259, 1993.

\bibitem{MW06}
A~J Majda and X~Wang.
\newblock {\em Nonlinear Dynamics and Statistical Theories for Basic
  Geophysical Flows}.
\newblock Cambridge University Press, 2006.

\bibitem{AM08}
R~Abramov and A~J Majda.
\newblock New approximations and tests of linear fluctuation-response for
  chaotic nonlinear forced-dissipative dynamical systems.
\newblock {\em J. Nonlinear Sci.}, 18:303, 2008.

\bibitem{N1}
R~V Abramov.
\newblock A theory of average response to large jump perturbations.
\newblock arXiv:1903.07226, 2019.

\bibitem{Seifert}
U~Seifert.
\newblock Stochastic thermodynamics, fluctuation theorems and molecular
  machines.
\newblock {\em Rep. Prog. Phys.}, 75:126001, 2012.

\bibitem{capklages}
G~Gradenigo, A~Puglisi, A~Sarracino, D~Villamaina, and A~Vulpiani.
\newblock Out-of-equilibrium generalized fluctuation-dissipation relations.
\newblock In C.~Jarzynski R.~Klages, W.~Just, editor, {\em Nonequilibrium
  Statistical Physics of Small Systems: Fluctuation Relations and Beyond}.
  Wiley-VCH, Weinheim, 2013.

\bibitem{CR03}
A~Crisanti and F~Ritort.
\newblock Violation of the fluctuation-dissipation theorem in glassy systems:
  basic notions and the numerical evidence.
\newblock {\em J. Phys. A}, 36:R181, 2003.

\bibitem{BPRV08}
U~Marini~Bettolo Marconi, A~Puglisi, L~Rondoni, and A~Vulpiani.
\newblock Fluctuation-dissipation: Response theory in statistical physics.
\newblock {\em Phys. Rep.}, 461:111, 2008.

\bibitem{cuglirev}
L~F Cugliandolo.
\newblock The effective temperature.
\newblock {\em J. Phys. A: Math. Theor.}, 44:483001, 2011.

\bibitem{maes3}
M~Baiesi and C~Maes.
\newblock An update on the nonequilibrium linear response.
\newblock {\em New J. Phys.}, 15:013004, 2013.

\bibitem{PSV17}
A~Puglisi, A~Sarracino, and A~Vulpiani.
\newblock Temperature in and out of equilibrium: A review of concepts, tools
  and attempts.
\newblock {\em Phys. Rep.}, 709-710:1, 2017.

\bibitem{baladi}
V~Baladi and D~Smania.
\newblock Linear response formula for piecewise expanding unimodal maps.
\newblock {\em Nonlinearity}, 21:677, 2008.

\bibitem{RS78}
R~H Rose and P~L Sulem.
\newblock Fully developed turbulence and statistical mechanics.
\newblock {\em J. {P}hys. ({P}aris)}, 39:441, 1978.

\bibitem{DH75}
U~Deker and F~Haake.
\newblock Fluctuation-dissipation theorems for classical processes.
\newblock {\em Phys. Rev. A}, 11:2043, 1975.

\bibitem{FIV90}
M~Falcioni, S~Isola, and A~Vulpiani.
\newblock Correlation functions and relaxation properties in chaotic dynamics
  and statistical mechanics.
\newblock {\em Physics {L}etters {A}}, 144:341, 1990.

\bibitem{BLMV03}
G~Boffetta, G~Lacorata, S~Musacchio, and A~Vulpiani.
\newblock Relaxation of finite perturbations: Beyond the fluctuation-response
  relation.
\newblock {\em Chaos}, 13:806, 2003.

\bibitem{vK71}
N~G van Kampen.
\newblock The case against linear response theory.
\newblock {\em Phys. {N}orv.}, 5:279, 1971.

\bibitem{K86}
R~Kubo.
\newblock Brownian motion and nonequilibrium statistical mechanics.
\newblock {\em Science}, 233:330, 1986.

\bibitem{Puglisi}
A~Puglisi.
\newblock {\em Transport and fluctuations in granular fluids: from Boltzmann
  equation to hydrodynamics, diffusion and motor effects}.
\newblock Springer, 2015.

\bibitem{RevModPhys.85.1143}
M~C Marchetti, J~F Joanny, S~Ramaswamy, T~B Liverpool, J~Prost, M~Rao, and
  R~Aditi Simha.
\newblock Hydrodynamics of soft active matter.
\newblock {\em Rev. Mod. Phys.}, 85:1143--1189, 2013.

\bibitem{K66}
R~Kubo.
\newblock The fluctuation-dissipation theorem.
\newblock {\em Rep. {P}rog. {P}hys.}, 29:255, 1966.

\bibitem{Joyce}
G~Joyce and D~Montgomery.
\newblock Negative temperature states for the two-dimensional guiding-centre
  plasma.
\newblock {\em J. Plasma Physics}, 10:107, 1973.

\bibitem{Braun}
S~Braun, J~P Ronzheimer, M~Schreiber, S~S Hodgman, T~Rom, I~Bloch, and
  U~Schneider.
\newblock Negative absolute temperature for motional degrees of freedom.
\newblock {\em Science}, 339:52, 2013.

\bibitem{N2}
F~Miceli, M~Baldovin, and A~Vulpiani.
\newblock Statistical mechanics of systems with long-range interactions and
  negative absolute temperature.
\newblock {\em Phys. Rev. E}, 99:042152, 2019.

\bibitem{N3}
D~Ruelle.
\newblock General linear response formula in statistical mechanics, and the
  fluctuation-dissipation theorem far from equilibrium.
\newblock {\em Phys. Lett. A}, 245:220, 1998.

\bibitem{CS07}
B~Cessac and J~A Sepulchre.
\newblock Linear response, susceptibility and resonance in chaotic toy models.
\newblock {\em Physica {D}}, 225:13, 2007.

\bibitem{N4}
M~Colangeli, L~Rondoni, and A~Vulpiani.
\newblock Fluctuation-dissipation relation for chaotic non-hamiltonian systems.
\newblock {\em J. Stat. Mech.}, 2012:L04002, 2012.

\bibitem{CM99}
C~Maes.
\newblock The fluctuation theorem as a {G}ibbs property.
\newblock {\em J. Stat. Phys.}, 95:367, 1999.

\bibitem{LCZ05}
E~Lippiello, F~Corberi, and M~Zannetti.
\newblock Off-equilibrium generalization of the fluctuation dissipation theorem
  for {I}sing spins and measurement of the linear response function.
\newblock {\em Phys. Rev. E}, 71:036104, 2005.

\bibitem{CLSZ10}
F~Corberi, E~Lippiello, A~Sarracino, and M~Zannetti.
\newblock Fluctuation-dissipation relations and field-free algorithms for the
  computation of response functions.
\newblock {\em Phys. Rev. E}, 81:011124, 2010.

\bibitem{Basu_2015}
U~Basu and C~Maes.
\newblock Nonequilibrium response and frenesy.
\newblock {\em Journal of Physics: Conference Series}, 638:012001, 2015.

\bibitem{PhysRevE.93.032128}
O~B\'enichou, P~Illien, G~Oshanin, A~Sarracino, and R~Voituriez.
\newblock Nonlinear response and emerging nonequilibrium microstructures for
  biased diffusion in confined crowded environments.
\newblock {\em Phys. Rev. E}, 93:032128, 2016.

\bibitem{baiesinew}
G~Teza, S~Iubini, M~Baiesi, A~L Stella, and C~Vanderzande.
\newblock Rate dependence of current and fluctuations in jump models with
  negative differential mobility.
\newblock arXiv:1904.05241, 2019.

\bibitem{BMW09}
M~Baiesi, C~Maes, and B~Wynants.
\newblock Fluctuations and response of nonequilibrium states.
\newblock {\em Phys. Rev. Lett.}, 103:010602, 2009.

\bibitem{maes1}
M~Baiesi, C~Maes, and B~Wynants.
\newblock Nonequilibrium linear response for markov dynamics, i: jump processes
  and overdamped diffusions.
\newblock {\em J. Stat. Phys.}, 137:1094, 2009.

\bibitem{maes2}
M~Baiesi, E~Boksenbojm, C~Maes, and B~Wynants.
\newblock Nonequilibrium linear response for markov dynamics, ii: Inertial
  dynamics.
\newblock {\em J. Stat. Phys.}, 139:492, 2010.

\bibitem{ss06}
T~Speck and U~Seifert.
\newblock Restoring a fluctuation-dissipation theorem in a nonequilibrium
  steady state.
\newblock {\em Europhys. Lett.}, 74:391, 2006.

\bibitem{Verley_2011}
G~Verley, K~Mallick, and D~Lacoste.
\newblock Modified fluctuation-dissipation theorem for non-equilibrium steady
  states and applications to molecular motors.
\newblock {\em {EPL} (Europhysics Letters)}, 93:10002, 2011.

\bibitem{PhysRevLett.103.090601}
J~Prost, J-F Joanny, and J~M~R Parrondo.
\newblock Generalized fluctuation-dissipation theorem for steady-state systems.
\newblock {\em Phys. Rev. Lett.}, 103:090601, 2009.

\bibitem{GPCCG09}
J~R Gomez-Solano, A~Petrosyan, S~Ciliberto, R~Chetrite, and K~Gawedzki.
\newblock Experimental verification of a modified fluctuation-dissipation
  relation for a micron-sized particle in a non-equilibrium steady state.
\newblock {\em Phys. Rev. Lett.}, 103:040601, 2009.

\bibitem{GPCM11}
J~R Gomez-Solano, A~Petrosyan, S~Ciliberto, and C~Maes.
\newblock Fluctuations and response in a non-equilibrium micron-sized system.
\newblock {\em J. Stat. Mech.}, page P01008, 2011.

\bibitem{CKP94}
L~F Cugliandolo, J~Kurchan, and G~Parisi.
\newblock Off equilibrium dynamics and aging in unfrustrated systems.
\newblock {\em J. Phys. I France}, 4:1641, 1994.

\bibitem{olla}
C~Landim, S~Olla, and S~R~S Varadhan.
\newblock On viscosity and fluctuation-dissipation in exclusion processes.
\newblock {\em J. Stat. Phys.}, 115:323, 2004.

\bibitem{falasco}
G~Falasco and M~Baiesi.
\newblock Nonequilibrium temperature response for stochastic overdamped
  systems.
\newblock {\em New Journal of Physics}, 18:043039, 2016.

\bibitem{falasco2}
G~Falasco and M~Baiesi.
\newblock Temperature response in nonequilibrium stochastic systems.
\newblock {\em Europhys. Lett.}, 113:20005, 2016.

\bibitem{bouchaud2005nonlinear}
J-P Bouchaud and G~Biroli.
\newblock Nonlinear susceptibility in glassy systems: A probe for cooperative
  dynamical length scales.
\newblock {\em Phys. Rev. B}, 72:064204, 2005.

\bibitem{LCSZ08a}
E~Lippiello, F~Corberi, A~Sarracino, and M~Zannetti.
\newblock Nonlinear susceptibilities and the measurement of a cooperative
  length.
\newblock {\em Phys. Rev. B}, 77:212201, 2008.

\bibitem{LCSZ08b}
E~Lippiello, F~Corberi, A~Sarracino, and M~Zannetti.
\newblock Nonlinear response and fluctuation-dissipation relations.
\newblock {\em Phys. Rev. E}, 78:041120, 2008.

\bibitem{crauste2010evidence}
C~Crauste-Thibierge, C~Brun, F~Ladieu, D~L’h{\^o}te, G~Biroli, and J-P
  Bouchaud.
\newblock Evidence of growing spatial correlations at the glass transition from
  nonlinear response experiments.
\newblock {\em Physical review letters}, 104:165703, 2010.

\bibitem{Basu}
U~Basu, M~Kr\"uger, A~Lazarescu, and C~Maes.
\newblock Frenetic aspects of second order response.
\newblock {\em Phys. Chem. Chem. Phys.}, 17:6653, 2015.

\bibitem{seifert05}
U~Seifert.
\newblock Entropy production along a stochastic trajectory and an integral
  fluctuation theorem.
\newblock {\em Phys. Rev. Lett.}, 95:040602, 2005.

\bibitem{maesbook}
C~Maes.
\newblock {\em Non-Dissipative Effects in Nonequilibrium Systems}.
\newblock Springer, 2018.

\bibitem{SVGP10}
A~Sarracino, D~Villamaina, G~Gradenigo, and A~Puglisi.
\newblock Irreversible dynamics of a massive intruder in dense granular fluids.
\newblock {\em Europhys. Lett.}, 92:34001, 2010.

\bibitem{SVCP10}
A~Sarracino, D~Villamaina, G~Costantini, and A~Puglisi.
\newblock Granular brownian motion.
\newblock {\em J. Stat. Mech.}, page P04013, 2010.

\bibitem{GPSV14}
A~Gnoli, A~Puglisi, A~Sarracino, and A~Vulpiani.
\newblock Nonequilibrium {B}rownian motion beyond the effective temperature.
\newblock {\em Plos One}, 9:e93720, 2014.

\bibitem{CPV12}
A~Crisanti, A~Puglisi, and D~Villamaina.
\newblock Non-equilibrium and information: the role of cross-correlations.
\newblock {\em Phys. Rev. E 85}, 85:061127, 2012.

\bibitem{CKP97}
L~F Cugliandolo, J~Kurchan, and L~Peliti.
\newblock Energy flow, partial equilibration, and effective temperatures in
  systems with slow dynamics.
\newblock {\em Phys. Rev. E}, 55:3898, 1997.

\bibitem{VBPV09}
D~Villamaina, A~Baldassarri, A~Puglisi, and A~Vulpiani.
\newblock Fluctuation dissipation relation: how does one compare correlation
  functions and responses?
\newblock {\em J. Stat. Mech.}, page P07024, 2009.

\bibitem{DMGBLN03}
G~D'Anna, P~Mayor, G~Gremaud, A~Barrat, V~Loreto, and F~Nori.
\newblock Observing brownian motion in vibration-fluidized granular matter.
\newblock {\em Nature}, 424:909, 2003.

\bibitem{BBDLMP05}
A~Baldassarri, A~Barrat, G~D'Anna, V~Loreto, P~Mayor, and A~Puglisi.
\newblock What is the temperature of a granular medium?
\newblock {\em Journal of Physics: Condensed Matter}, 17:S2405, 2005.

\bibitem{keys2007measurement}
A~S Keys, A~R Abate, S~C Glotzer, and D~J Durian.
\newblock Measurement of growing dynamical length scales and prediction of the
  jamming transition in a granular material.
\newblock {\em Nature physics}, 3:260, 2007.

\bibitem{PhysRevE.77.051111}
D~Loi, S~Mossa, and L~F Cugliandolo.
\newblock Effective temperature of active matter.
\newblock {\em Phys. Rev. E}, 77:051111, 2008.

\bibitem{Wang15184}
S~Wang and P~G Wolynes.
\newblock On the spontaneous collective motion of active matter.
\newblock {\em Proceedings of the National Academy of Sciences},
  108:15184--15189, 2011.

\bibitem{PhysRevE.90.012111}
G~Szamel.
\newblock Self-propelled particle in an external potential: Existence of an
  effective temperature.
\newblock {\em Phys. Rev. E}, 90:012111, 2014.

\bibitem{Levis_2015}
D~Levis and L~Berthier.
\newblock From single-particle to collective effective temperatures in an
  active fluid of self-propelled particles.
\newblock {\em {EPL} (Europhysics Letters)}, 111:60006, 2015.

\bibitem{Han7513}
M~Han, J~Yan, S~Granick, and E~Luijten.
\newblock Effective temperature concept evaluated in an active colloid mixture.
\newblock {\em Proceedings of the National Academy of Sciences},
  114:7513--7518, 2017.

\bibitem{PhysRevE.97.032125}
C~M Rohwer, A~Solon, M~Kardar, and M~Kr\"uger.
\newblock Nonequilibrium forces following quenches in active and thermal
  matter.
\newblock {\em Phys. Rev. E}, 97:032125, 2018.

\bibitem{dieterich2015single}
E~Dieterich, J~Camunas-Soler, M~Ribezzi-Crivellari, U~Seifert, and F~Ritort.
\newblock Single-molecule measurement of the effective temperature in
  non-equilibrium steady states.
\newblock {\em Nature Physics}, 11:971, 2015.

\bibitem{preisler2016configurational}
Z~Preisler and M~Dijkstra.
\newblock Configurational entropy and effective temperature in systems of
  active brownian particles.
\newblock {\em Soft matter}, 12:6043--6048, 2016.

\bibitem{PhysRevLett.118.015702}
C~M Rohwer, M~Kardar, and M~Kr\"uger.
\newblock Transient casimir forces from quenches in thermal and active matter.
\newblock {\em Phys. Rev. Lett.}, 118:015702, 2017.

\bibitem{workamp2018symmetry}
M~Workamp, G~Ramirez, K~E Daniels, and J~A Dijksman.
\newblock Symmetry-reversals in chiral active matter.
\newblock {\em Soft matter}, 14:5572--5580, 2018.

\bibitem{seifert2019stochastic}
U~Seifert.
\newblock From stochastic thermodynamics to thermodynamic inference.
\newblock {\em Annual Review of Condensed Matter Physics}, 10:171--192, 2019.

\bibitem{cugliandolo2019effective}
L~F Cugliandolo, G~Gonnella, and I~Petrelli.
\newblock Effective temperature in active brownian particles.
\newblock {\em Fluctuation and Noise Letters}, 18:1940008, 2019.

\bibitem{golestanian}
R~Golestanian.
\newblock Bose–{E}instein condensation in scalar active matter with
  diffusivity edge.
\newblock {\em Phys. Rev. E}, 100:010601, 2019.

\bibitem{klongvessa}
N~Klongvessa, F~Ginot, C~Ybert, C~Cottin-Bizonne, and M~Leocmach.
\newblock Active glass: ergodicity breaking dramatically affects response to
  self-propulsion.
\newblock arXiv:1902.01746, 2019.

\bibitem{durian1}
I~K Ono, C~S O'Hern, D~J Durian, S~A Langer, A~J Liu, and S~R Nagel.
\newblock Effective temperatures of a driven system near jamming.
\newblock {\em Phys. Rev. Lett.}, 89:095703, 2002.

\bibitem{durian2}
R~P Ojha, P~A Lemieux, P~K Dixon, A~J Liu, and D~J Durian.
\newblock Statistical mechanics of a gas-fluidized particle.
\newblock {\em Nature}, 427:521, 2004.

\bibitem{durian3}
A~R Abate and D~J Durian.
\newblock Effective temperatures and activated dynamics for a two-dimensional
  air-driven granular system on two approaches to jamming.
\newblock {\em Phys. Rev. Lett.}, 101:245701, 2008.

\bibitem{BG90}
J~P Bouchaud and A~Georges.
\newblock Anomalous diffusion in disordered media: Statistical mechanisms,
  models and physical applications.
\newblock {\em Phys. Rep.}, 195:127, 1990.

\bibitem{GSGWS96}
Q~Gu, E~A Schiff, S~Grebner, F~Wang, and R~Schwarz.
\newblock Non-gaussian transport measurements and the {E}instein relation in
  amorphous silicon.
\newblock {\em Phys. Rev. Lett.}, 76:3196, 1996.

\bibitem{CMMV99}
P~Castiglione, A~Mazzino, P~Muratore-Ginanneschi, and A~Vulpiani.
\newblock On strong anomalous diffusion.
\newblock {\em Physica D}, 134:75, 1999.

\bibitem{MK00}
R~Metzler and J~Klafter.
\newblock The random walk's guide to anomalous diffusion: a fractional dynamics
  approach.
\newblock {\em Phys. Rep.}, 339:1, 2000.

\bibitem{BC05}
R~Burioni and D~Cassi.
\newblock Random walks on graphs: ideas, techniques and results.
\newblock {\em J. Phys. A:Math. Gen.}, 38:R45, 2005.

\bibitem{BO02}
O~B\'enichou and G~Oshanin.
\newblock Ultraslow vacancy-mediated tracer diffusion in two dimensions: The
  einstein relation verified.
\newblock {\em Phys. Rev. E}, 66:031101, 2002.

\bibitem{MBK99}
R~Metzler, E~Barkai, and J~Klafter.
\newblock Anomalous diffusion and relaxation close to thermal equilibrium: A
  fractional {F}okker-{P}lanck equation approach.
\newblock {\em Phys. Rev. Lett}, 82:3563, 1999.

\bibitem{BF98}
E~Barkai and V~N Fleurov.
\newblock Generalized {E}instein relation: A stochastic modeling approach.
\newblock {\em Phys. Rev. E}, 58:1296, 1998.

\bibitem{CK09}
A~V Chechkin and R~Klages.
\newblock Fluctuation relations for anomalous dynamics.
\newblock {\em J. Stat. Mech.}, page L03002, 2009.

\bibitem{LATBL10}
L~Lizana, T~Ambj{\"o}rnsson, A~Taloni, E~Barkai, and M~A Lomholt.
\newblock Foundation of fractional {L}angevin equation: {H}armonization of a
  many-body problem.
\newblock {\em Phys. Rev. E}, 81:51118, 2010.

\bibitem{VPV08}
D~Villamaina, A~Puglisi, and A~Vulpiani.
\newblock The fluctuation-dissipation relation in sub-diffusive systems: the
  case of granular single-file diffusion.
\newblock {\em J. Stat. Mech.}, page L10001, 2008.

\bibitem{HKK96}
K~Hahn, J~K{\"a}rger, and V~Kukla.
\newblock Single-file diffusion observation.
\newblock {\em Phys. Rev. Lett.}, 76:2762, 1996.

\bibitem{comb}
D~Villamaina, A~Sarracino, G~Gradenigo, A~Puglisi, and A~Vulpiani.
\newblock On anomalous diffusion and the out of equilibrium response function
  in one-dimensional models.
\newblock {\em J. Stat. Mech.}, 2011:L01002, 2011.

\bibitem{Forte_2013}
G~Forte, R~Burioni, F~Cecconi, and A~Vulpiani.
\newblock Anomalous diffusion and response in branched systems: a simple
  analysis.
\newblock {\em Journal of Physics: Condensed Matter}, 25:465106, 2013.

\bibitem{Gradenigo_2012}
G~Gradenigo, A~Sarracino, D~Villamaina, and A~Vulpiani.
\newblock Einstein relation in superdiffusive systems.
\newblock {\em Journal of Statistical Mechanics: Theory and Experiment},
  2012:L06001, 2012.

\bibitem{PhysRevE.87.030104}
D~Froemberg and E~Barkai.
\newblock Time-averaged einstein relation and fluctuating diffusivities for the
  l\'evy walk.
\newblock {\em Phys. Rev. E}, 87:030104, 2013.

\bibitem{godec2013linear}
A~Godec and R~Metzler.
\newblock Linear response, fluctuation-dissipation, and finite-system-size
  effects in superdiffusion.
\newblock {\em Physical Review E}, 88:012116, 2013.

\bibitem{kusmierz2018thermodynamics}
L~Ku{\'s}mierz, B~Dybiec, and E~Gudowska-Nowak.
\newblock Thermodynamics of superdiffusion generated by l{\'e}vy--wiener
  fluctuating forces.
\newblock {\em Entropy}, 20:658, 2018.

\bibitem{costa2003fluctuation}
I~V~L Costa, R~Morgado, M~V B~T Lima, and F~A Oliveira.
\newblock The fluctuation-dissipation theorem fails for fast superdiffusion.
\newblock {\em EPL (Europhysics Letters)}, 63:173, 2003.

\bibitem{dieterich2015fluctuation}
P~Dieterich, R~Klages, and A~V Chechkin.
\newblock Fluctuation relations for anomalous dynamics generated by
  time-fractional fokker--planck equations.
\newblock {\em New Journal of Physics}, 17:075004, 2015.

\bibitem{pottier}
N~Pottier and A~Mauger.
\newblock Anomalous diffusion of a particle in an aging medium.
\newblock {\em Physica A}, 332:15, 2004.

\bibitem{pottier1}
N~Pottier.
\newblock Aging properties of an anomalouslydi"using particule.
\newblock {\em Physica A}, 317:371, 2003.

\bibitem{PhysRevLett.115.220601}
O~B\'enichou, P~Illien, G~Oshanin, A~Sarracino, and R~Voituriez.
\newblock Diffusion and subdiffusion of interacting particles on comblike
  structures.
\newblock {\em Phys. Rev. Lett.}, 115:220601, 2015.

\bibitem{PhysRevLett.117.174501}
A~Sarracino, F~Cecconi, A~Puglisi, and A~Vulpiani.
\newblock Nonlinear response of inertial tracers in steady laminar flows:
  Differential and absolute negative mobility.
\newblock {\em Phys. Rev. Lett.}, 117:174501, 2016.

\bibitem{N5}
G~Lacorata and A~Vulpiani.
\newblock Fluctuation-response relation and modeling in systems with fast and
  slow dynamics.
\newblock {\em Nonlinear Proc. in Geophys.}, 14:681, 2007.

\bibitem{Wormell}
C~L Wormell and G~A Gottwald.
\newblock On the validity of linear response theory in high-dimensional
  deterministic dynamical systems.
\newblock {\em J. Stat. Phys.}, 172:1479, 2018.

\bibitem{DEMAGISTRIS201565}
G~De Magistris and D~Marenduzzo.
\newblock An introduction to the physics of active matter.
\newblock {\em Physica A: Statistical Mechanics and its Applications}, 418:65
  -- 77, 2015.
\newblock Proceedings of the 13th International Summer School on Fundamental
  Problems in Statistical Physics.

\bibitem{PhysRevLett.117.038103}
E~Fodor, C~Nardini, M~E Cates, J~Tailleur, P~Visco, and F~van Wijland.
\newblock How far from equilibrium is active matter?
\newblock {\em Phys. Rev. Lett.}, 117:038103, 2016.

\bibitem{PhysRevLett.119.258001}
D~Mandal, K~Klymko, and M~R DeWeese.
\newblock Entropy production and fluctuation theorems for active matter.
\newblock {\em Phys. Rev. Lett.}, 119:258001, 2017.

\bibitem{PhysRevE.98.020604}
S~Shankar and M~C Marchetti.
\newblock Hidden entropy production and work fluctuations in an ideal active
  gas.
\newblock {\em Phys. Rev. E}, 98:020604, 2018.

\bibitem{e19070356}
A~Puglisi and U~Marini~Bettolo Marconi.
\newblock Clausius relation for active particles: What can we learn from
  fluctuations.
\newblock {\em Entropy}, 19, 2017.

\bibitem{PhysRevE.90.052130}
A~Suma, G~Gonnella, G~Laghezza, A~Lamura, A~Mossa, and L~F Cugliandolo.
\newblock Dynamics of a homogeneous active dumbbell system.
\newblock {\em Phys. Rev. E}, 90:052130, 2014.

\bibitem{maggi}
C~Maggi, M~Paoluzzi, L~Angelani, and R~Di Leonardo.
\newblock Memory-less response and violation of the fluctuation-dissipation
  theorem in colloids suspended in an active bath.
\newblock {\em Sci. Rep.}, 7:17588, 2017.

\bibitem{Szamel_2017}
G~Szamel.
\newblock Evaluating linear response in active systems with no perturbing
  field.
\newblock {\em {EPL} (Europhysics Letters)}, 117:50010, 2017.

\bibitem{chong}
C~Shen and H~D Ou-Yang.
\newblock Fluctuation-dissipation of an active brownian particle under
  confinement.
\newblock In {\em Optical Trapping and Optical Micromanipulation XV}, volume
  10723. International Society for Optics and Photonics, 2018, 2018.

\bibitem{burk}
E~W Burkholder and J~F Brady.
\newblock Fluctuation-dissipation in active matter.
\newblock {\em J. Chem. Phys.}, 150:184901, 2019.

\bibitem{pagona}
S~Dal Cengio, D~Levis, and I~Pagonabarraga.
\newblock Linear response theory and {G}reen-{K}ubo relations for active
  matter.
\newblock arXiv:1907.02560, 2019.

\bibitem{maggi1}
C~Maggi, U~Marini~Bettolo Marconi, N~Gnan, and R~Di Leonardo.
\newblock Multidimensional stationary probability distribution for interacting
  active particles.
\newblock {\em Sci. Rep.}, 5:10742, 2015.

\bibitem{Caprini_2018}
L~Caprini, U~Marini~Bettolo Marconi, and A~Vulpiani.
\newblock Linear response and correlation of a self-propelled particle in the
  presence of external fields.
\newblock {\em Journal of Statistical Mechanics: Theory and Experiment},
  2018:033203, 2018.

\bibitem{safaverdi}
C~Maes, S~Safaverdi, P~Visco, and F~van Wijland.
\newblock Fluctuation-response relations for nonequilibrium diffusions with
  memory.
\newblock {\em Phys. Rev. E}, 87:022125, 2013.

\bibitem{PhysRevLett.80.5044}
F~Schweitzer, W~Ebeling, and B~Tilch.
\newblock Complex motion of brownian particles with energy depots.
\newblock {\em Phys. Rev. Lett.}, 80:5044--5047, 1998.

\bibitem{strutt1945theory}
J~W Strutt.
\newblock {\em The theory of sound}.
\newblock Dover, 1945.

\bibitem{PhysRevE.88.052124}
A~Sarracino.
\newblock Time asymmetry of the kramers equation with nonlinear friction:
  Fluctuation-dissipation relation and ratchet effect.
\newblock {\em Phys. Rev. E}, 88:052124, 2013.

\end{thebibliography}
\bibliographystyle{unsrt}

\end{document}